\documentclass[aps,twocolumn,prl,showpacs,superscriptaddress]{revtex4}

\usepackage{graphicx}

\def\tmc{T_\mathrm{MC}}
\def\te{T_\mathrm{eff}}

\begin{document}

\title{Asymptotic aging in structural glasses}

\author{T.~S.~Grigera}
\altaffiliation{Present address: Instituto de Investigaciones
Fisicoqu\'\i{}micas Te\'oricas y Aplicadas (INIFTA, CONICET--UNLP),
c.c. 16, suc. 4, 1900 La Plata, Argentina.}
\affiliation{Centro Studi e Ricerche ``Enrico Fermi'', via Panisperna,
89A, 00184 Roma, Italy}

\author{V.~Mart\'\i{}n-Mayor} 
\affiliation{Departamento de F\'\i{}sica Te\'orica, Universidad
Complutense, Avenida Complutense, 28040 Madrid, Spain}
\affiliation{Instituto de Biocomputaci\'on y F\'{\i}sica de Sistemas
Complejos (BIFI). Universidad de Zaragoza, 50009 Zaragoza, Spain.}

\author{G.~Parisi} 
\affiliation{Dipartimento di Fisica, Universit\`a di Roma ``La
Sapienza'', P.le Aldo Moro 2, 00185 Roma, Italy- SMC-INFM unit\`a di
Roma - Italy INFN sezione di Roma}

\author{P.~Verrocchio}
\affiliation{Departamento de F\'\i{}sica Te\'orica, Universidad
Complutense, Avenida Complutense, 28040 Madrid, Spain}
\affiliation{Instituto de Biocomputaci\'on y F\'{\i}sica de Sistemas
Complejos (BIFI). Universidad de Zaragoza, 50009 Zaragoza, Spain.}

\date{April 6, 2004}

\begin{abstract}
Using a non-local Monte Carlo algorithm, we study the aging of a
fragile glass, being able to follow it up to equilibibrium down to
$0.89 \tmc$ ($\tmc$ is the Mode-Coupling temperature) and up to
unprecedentedly large waiting times at lower temperatures. We show
that the fluctuation-dissipation ratio is independent of the dynamics
chosen and is compatible with a phase transition, and that the scaling
behaviour of the aging part of the correlation supports the full-aging
scenario.
\end{abstract}

\maketitle

{\em Aging} is found in many complex systems out of equilibrium, like
supercooled liquids \cite{liquids}, polymers \cite{Struick78},
colloids \cite{colloids}, or spin-glasses \cite{sg}, and understanding
it is a necessary step towards a unified description of such
systems~\cite{DYNBOOK,AGINGKURCHAN}.  After a short transient since
preparation, a state is reached in which one-time observables (e.g.\
energy, enthalpy) vary extremely slowly, while two-time quantities
(correlations, susceptibilities) strongly depend on the {\em age\/}
(or {\em waiting time\/} $t_w$, i.e.\ the time elapsed since
preparation) of the system as well as on frequency $\omega$ (or the
measurement time $t$). Despite recent efforts, our knowledge of aging
of real materials is scant in the theoretically important regime of
large $t_w$ and small frequency, where universal features should show
up \cite{DYNBOOK}. Two issues still open are the scaling of
correlations and the behavior of the fluctuation-dissipation ratio.

Consider observables $A$ and $B$ ($B$ couples to an external field
$h$). The susceptibility $\chi$ (i.e.\ the time integral of the linear
response $R(t_w,t+t_w)\equiv \delta \langle A(t+tw) \rangle/\delta
h(t_w)|_{h=0}$) and the correlation function
$C(t_w,t+t_w)\equiv\langle A(t+t_w)B(t_w)\rangle$ are expected to be
of the form~\cite{DYNBOOK}
\begin{equation}
C(t_w,t_w+t) = C_\mathrm{st}(t) + C_\mathrm{ag}\left(
\frac{g(t_w+t)}{g(t_w)} \right) ,   \label{CAG}
\label{scaling}
\end{equation}
where $g(t)$ is a monotonic function acting as an 'effective'
correlation time, and $C_\mathrm{ag}$ describes the aging of the
system~\cite{Vincent97}.  Most published studies focus on the scaling
properties of $C_\mathrm{ag}$: it is generally a function of
$t/t_w^\mu$, but there is a lack of universality in the values of the
exponent $\mu$, embarrassing in view of the claimed equivalence of
complex systems. \emph{Full aging} ($\mu=1$) has been clearly observed
so far only in spin-glasses \cite{agingsg}. For colloids, both
\emph{superaging} ($\mu>1$ \cite{Bouchaud99}) and full aging has been
reported \cite{colloids-ag}. Polymers show rather \emph{subaging}
($\mu<1$) \cite{Struick78,Ciliberto03}, as has also been observed in
simple liquids \cite{Kob97}. However, the values quoted often
correspond to different time regimes, and the regime where
$t_w\to\infty$ with $t/t_w$ fixed has not been carefully studied
(except for spin glasses). For example in glycerol \cite{Nagel98} full
aging has not been seen either close to the glass temperature, $T_g$
(almost at equilibrium) or at lower temperatures $T$.  In both
regimes the explored frequencies were much larger than $1/t_w$.

Aging is also characterized by a non-trivial behavior of the
fluctuation-dissipation ratio (FDR), namely
\begin{equation}
 X(t_w,t+t_w) = { T R(t_w,t+t_w) \over  dC(t_w,t_w+t)/dt_w }.
\label{FDR}
\end{equation}
The fluctuation-dissipation theorem (FDT) states that $X=1$ in
thermodynamic equilibrium, but this need not be so during aging, and
\emph{FDT violations} (i.e.\ $X\neq1$) are observed.
Experiments~\cite{FDTexpliq,FDTexpnoliq}, mean-field results
\cite{CuKuPe} and simulations ~\cite{FDTnumliq,Sciortino01} suggest
that the FDR depends on time only through the correlation function,
i.e.\ $X=X[C(t_w,t_w+t)]$. In structural glasses, in which we
concentrate from now on, simulations also show that at fixed $t_w$,
$X$ takes essentially two values: $X(C)=1$ for $C$ greater than some
$q_\mathrm{EA}(T)$ (called Edwards-Anderson parameter) and
$X(C)=x(t_w) <1$ for $C<q_\mathrm{EA}(T)$.  Since $T/X$ can be
interpreted as an effective temperature $\te$ \cite{CuKuPe}, it seems
that FDT violations in structural glasses can be characterized by a
single time-dependent $\te(t_w) \equiv T/x(t_w)$, related to the
slowest degrees of freedom. This lacks experimental confirmation.
(Note that other definitions of effective temperautres have been
explored \cite{Nagel98,fictive}).  Also open is the issue of the
behavior of $\te(t_w)$ as $t_w\to\infty$ (numerical data available
cover only very short waiting times in the sense that one-time
quantities are still fastly evolving \cite{Kob97,Sciortino01}), of
great theoretical interest because it is related to the possible {\em
thermodynamic} meaning of $\te$ \cite{CuKuPe}.

In this paper we study the aging dynamics down to $0.53\,\tmc$ ($\tmc$
is the Mode-Coupling \cite{Goetze92} temperature, below which dynamics
slows down dramatically), reaching very large waiting times. This can
be achieved through the use of a non-local algorithm (SMC
\cite{SWAP}), which greatly accelerates the dynamics.  We reach an
asymptotic regime where the correlation function shows full aging
within errors (supporting the analogy with spin glasses
\cite{agingsg}), and where FDT violations are independent of the
dynamics and of the age of the system.

We have simulated the soft-sphere binary mixture~\cite{ss} (pair
potential $V_{AB}(r)=(\sigma_{AB}/r)^{12}$, diameter ratio 1.2), a
simple fragile glassformer, using a {\em non-local} Metropolis Monte
Carlo algorithm (hereafter SMC) \cite{SWAP} which adds swap moves
(with probability $p$) to standard {\em local} Monte Carlo
(LMC). Although swap acceptance is very low ($\approx 3\cdot 10^{-3}$)
the equilibration time is considerably shortened; e.g.\ at
$0.89\,\tmc$ extrapolations estimate it to be three orders of
magnitude larger for LMC than for SMC (note that other non-local
algorithms have proved useful in simulations of structural
glasses~\cite{Santen99}). We used the following protocol: Starting
from a random configuration, a system of $N=2048$ particles was
instantaneously quenched to the final temperature $T$, and let evolve
for $t_w$ steps. This preparation was done with the SMC algorithm with
$p=0.1$, which gives the faster equilibration for this system
size. After $t_w$, the correlation and response functions in the
presence of an external field $h$ were computed, mostly in SMC runs
with $p=0.1$, but also in LMC and SMC runs with different $p$ in order
to assess the dependence of the results on the dynamics. Due to the
swap moves, particle diffusion is not a convenient
observable. Instead, we divided the simulation box in $N_c$ cubic
subcells and considered the quantity
\begin{equation}
A(t)=\frac{1}{N} \sum_{\alpha=1}^{N_c} \epsilon_\alpha n_\alpha(t),
\label{observable}
\end{equation}
where $\epsilon_\alpha=\pm 1$ randomly and $n_\alpha$ is the
occupation number of subcell $\alpha$. The side of the subcells was
about $0.35 \, \sigma_{AA}$ so that essentially $n_\alpha=0,1$. Note that
swap moves do not change $A(t)$.  To measure response, a term $\lambda
N A$ was added to the Hamiltonian, with $\lambda \equiv h k_B T$ ($h$
is dimensionless).  We considered the correlation $C(t_w,t_w+t) \equiv
\overline{\langle N A(t_w) A(t_w+t) \rangle}$, where
$\overline{\langle \ldots \rangle}$ means average over both thermal
histories and the $\epsilon_\alpha$, together with the integrated
response $k_B T \chi(t_w,t_w+t) \equiv \overline{\langle A(t_w+t)
\rangle /h}$ \cite{cm}.

\begin{figure}%[h!]
\includegraphics[angle=270,width=\columnwidth]{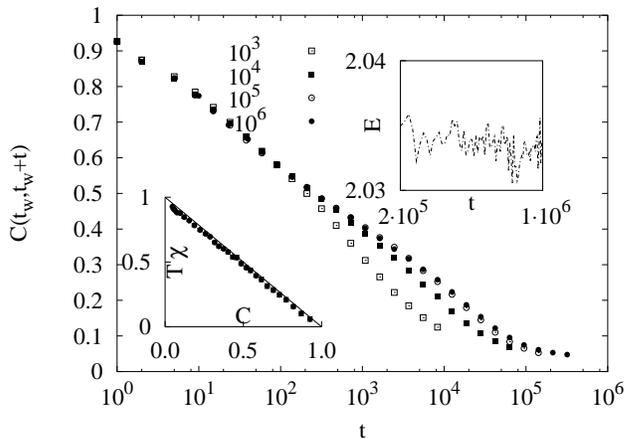}
\caption{Correlation function $C(t_w,t_w+t)$ vs.\ $t$ for $T=0.89
\tmc$ at $t_w=10^3,10^4,10^5,10^6$ (24 samples). Bottom, left:
integrated response $T \chi$ vs.\ correlation function $C$ at
$t_w=10^6$. Top, right: Energy per particle $E$ vs.\ $t$ during a SMC
quench with $p=0.1$. Error bars are of the order of point size.}
\label{equilibrio}
\end{figure}

With SMC we can equilibrate the system down to $T=0.89\,\tmc$. The
correlation $C(t_w,t_w+t)$ shows aging up to $t_w=10^5$, but does not
change between $t_w=10^5$ and $10^6$, which is approximately the
region where the energy reaches a stationary value
(Fig.~\ref{equilibrio}). We conservatively estimate the
autocorrelation time as the time $\tau$ needed for $C$ to reach the
asymptotic value $N/N_c$ ($\sim 0.04$), obtaining $\tau=2\times 10^5$,
much smaller than $10^6$ (the total lenght of the simulation). Hence
we claim that the system has equilibrated, which is further confirmed
by the fact that the FDT holds. In contrast, well below $0.89 \,\tmc$
the system is out of equilibrium up to $t_w = 2 \times 10^7$ (our
largest observational time).  A stretched exponential fit of the
equilibrium correlation in the late $\alpha$-relaxation regime yields
a stretching exponent $\beta \sim 0.3$.  The equilibrium LMC
correlation function does not decay to $N/Nc$ within the simulated
times, hence it is still an open point whether SMC changes the shape
of the correlations in equilibrium, or whether the two dynamics are
related by a simple rescaling of time.

\begin{figure}%[h!]
\includegraphics[angle=270,width=\columnwidth]{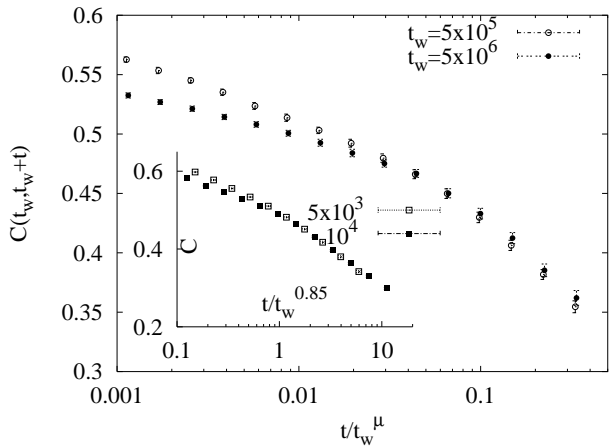}
\caption{$C$ vs.\ $t/t_w^{\mu}$, $\mu = 1.05(6)$, for $t_w=5\times
10^5, 5\times 10^6$ at $T=0.53 \: \tmc$ from SMC runs ($24$
samples). Inset: $C$ vs.\ $t/t_w^{0.85}$ for $t_w=5\times 10^3, 10^4$
($48$ samples).}
\label{full}
\end{figure}

We first address the issue of the scaling of the correlation during
aging at $T=0.53\, \tmc$ (in general far below $T_g$, e.g.\ for
glycerol this corresponds to $T\sim 140\,$K, while $T_g \sim
190\,$K). With SMC we find (Fig.~\ref{full}) that the correlations for
$t_w=5\times10^5,\,5\times 10^6$ can be made to collapse by plotting
them as a function of $t/t_w^{\mu}$ with $\mu = 1.05(6)$, compatible
with full aging. The collapse applies to the aging part
($C_\mathrm{ag}$, eq.~\ref{CAG}), which dominates the correlation for
$t/t_w > 0.1$ ($\omega t_w < 10$), as has also been observed in spin
glasses \cite{agingsg}. The two shortest $t_w$'s (inset) can instead
be scaled with $\mu \sim 0.85$. The same value (within errors) was
found in molecular dynamics simulations of the Lennard-Jones binary
mixture \cite{Kob97}, so we argue that the accelerated dynamics does
not affect the scaling. If one insists on scaling all curves, it can
be done reasonably well using $\mu \sim 0.9$, though this is likely an
artifact of mixing two different regimes. The relevant point is that
$\mu \sim 1$ is seen clearly only for $t_w \gg 1$ and in the $t \sim
t_w$ region, which is where it is expected to hold \cite{Bouchaud99},
if structural glasses share the dynamic properties of spin-glasses
\cite{DYNBOOK}. The failure of full aging for $t/t_w \ll 1$ is hence
in agreement with dielectric susceptibility measurements in
glycerol~\cite{Nagel98}. We are not aware of experimental studies in
the conditions where we find full aging, but such measurements are
clearly needed.

\begin{figure}%[h!]
\includegraphics[angle=270,width=\columnwidth]{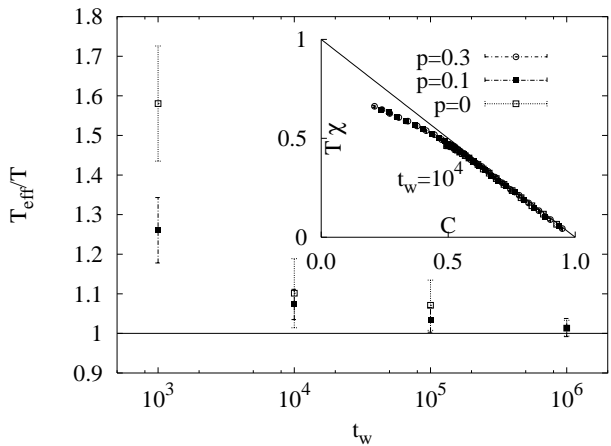}
\caption{$\te/T$ vs.\ $t_w$ for SMC and LMC runs at $T=0.89\, \tmc$
(16 samples), obtained by a linear fit of the points of the parametric
$T \chi$ vs. $C$ plots deviating from the FDT line. Errors were
estimated with the jacknife method \cite{Jacknife}. Inset: $T \chi$
vs.$C$ for $t_w=10^4$ at $T=0.53\,\tmc$ for $p=0$ (LMC) and $p=0.1,
0.3$ (SMC), $N=20000$ (8 samples).}
\label{indipendenza}
\end{figure}

A second important result is that although the susceptibility and
correlation are affected by the choice of dynamics, the FDR is not.
In fact, Fig.~\ref{indipendenza} shows the ratio $\te/T$ (i.e.\ the
inverse of the FDR) at $T= 0.89\, \tmc$ during aging and up to
equilibration for both SMC and LMC algorithms, obtained measuring the
FDR in simulations that used configurations taken along the SMC quench
as a starting point. After a short transient ($\sim 10^4$ steps) the
FDRs become indistinguishable within errors. At $T=0.53\,\tmc$ and
with LMC, we can reach the region of FDT violations only for
$t_w=10^4$, so we look at the FDR at fixed $t_w$ for LMC and SMC with
$p=0.1$ and $0.3$, obtaining a good agreement
(Fig.~\ref{indipendenza}, inset).

\begin{figure}%[h!]
\includegraphics[angle=270,width=\columnwidth]{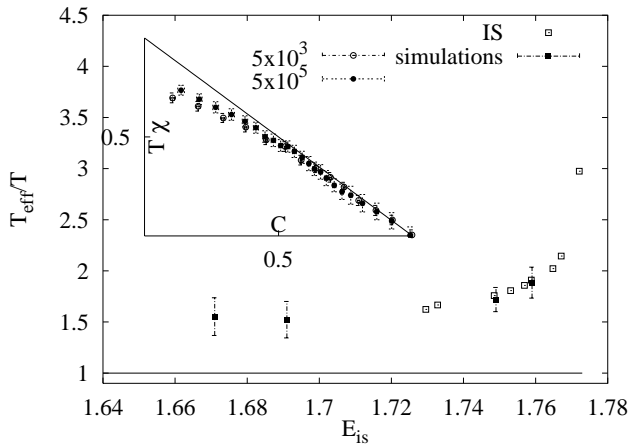}
\caption{$\te/T$ for $T=0.53 \tmc$ vs. the the istantaneous IS energy
$E_{IS}(t_w)$ measured in SMC runs. $E_{is}(t_w)$ is found by energy
minimization starting from $50$ instantaneous configurations within a
small time window ($<<t_w$) around $t_w$. We also show the $\te$
predicted by the IS approach ~\cite{Sciortino01}.  Inset: the $T \chi$
vs.$C$ plot for two different $t_w$.}
\label{transizione}
\end{figure}

Finally, we investigate the FDR for large times at $T=0.53\, \tmc$.
In Fig.~\ref{transizione} we plot $\te$ at $t_w=5\times 10^3,$ $10^4,$
$5\times 10^5$ and $5\times 10^6$ as a function of the instantaneous
inherent structure (IS) energy $E_\mathrm{IS}(t_w)$. We also plot
$\te$ computed according to the IS approach~\cite{Sciortino01},
$\te^{-1} = {\partial \Sigma}/{\partial f}$, where $\Sigma(f)$ is the
logarithm of the number of IS with free-energy $f$, and ${\partial
\Sigma}/{\partial f}$ is obtained as in
ref.~\onlinecite{Sciortino01}).  This idea (which makes no prediction
about the $t_w\to\infty$ limit of $\te$) had previously been confirmed
only in the very early aging regime by molecular dynamic simulations
\cite{Sciortino01}. Our results show a reasonable agreement even at
quite large times.

The limiting value of $\te$ as $t_w\to\infty$ is of great theoretical
interest. If the system eventually equilibrates, then $\te\to T$, as
we have found for $T=0.89 \,\tmc$. Approaches that consider aging a
result of critical slowing down due to the proximity of a critical
point which is never reached (beacuse it is located at
$T=0$~\cite{zero}, or because of the impossibility to establish a
``liquid'' long range order~\cite{tarjus}) predict this to be the case
for all temperatures. A different view relates the asymptotic value of
the FDR to a thermodynamic transition described by replica symmetry
breaking \cite{trans}. Above the transition, $X(C)$ is predicted to
reach slowly the equilibrium value $1$ (so $\te\to
T$~\cite{AGINGKURCHAN}), while below the FDR should remain non trivial
and $\te$ tend to a constant $>T$, since the system never
equilibrates. In this scenario the asymptotic FDR is claimed to
classify complex systems in universality classes \cite{DYNBOOK,trans}.
A third possibility is that FDT violations are due to nucleation and
slow growth of the crystal phase \cite{crystal}, in which case at long
times one expects the coarsening regime to be reached, and so
$\te\to\infty$.

Our results for $0.53\,\tmc$ do not seem to support this last
possibility. The data are instead compatible with the presence of a
thermodynamic replica symmetry breaking (RSB) transition \cite{trans},
since FDR does not seem to change between $t_w=5 \times 10^5$ and
$t_w=5 \times 10^6$ ($E_{IS}$ are respectively $1.691$ and
$1.671$). Note that this is the same regime where the system displays
full aging.  It cannot be excluded that $\te\to T$, but it looks less
likely if we note that extrapolating $E_\mathrm{IS}(t_w)$ to $t_w \to
\infty$ with a power-law gives an asymptotic $E_\mathrm{IS}=1.642$.
In the first approximation the RSB approach predicts that $\te$ equals
the transition temperature, which unfortunately has been only roughly
estimated\cite{trans}.  We just observe that, at the qualitative
level, the fact that the measured $\te/T$ in Fig.~\ref{transizione}
levels off at a value greater than 1 in the late aging regime supports
the RSB scenario.

In summary, we have for the first time studied numerically the late
aging regime of a simple glass-forming liquid using local and
non-local Monte Carlo (SMC). We find that the scaling of the
correlation functions and the FDR during aging do not depend on the
dynamics. This is a strong generalization of the previous finding
\cite{kob-indep} that equilibrium relaxation in the Lennard-Jones
mixture is qualitatively identical under different \emph{local}
dynamics (except, as here, for very short times). We have found that
correlation functions in the late aging regime show within errors
full-aging scaling, suggesting an equivalence between the aging
dynamics of structural and spin glasses. This should be searched
experimentally at frequencies comparable or shorter than $1/t_w$. We
also measured the FDR while taking one-time quantities closer to
assimptotic values than in previous studies. FDT violations do not
imply a thermodynamic transition.  However, if a transition does
exist, there should be a correspondence between the asympotic $\te$,
which is accessible to experiments, and the order parameter, which is
not \cite{FDTequivalence}. The FDRs measured in experiments
\cite{FDTexpliq,FDTexpnoliq} and simulations \cite{Sciortino01} up to
now depend strongly on the age of the system, hence their utility in
investigating the existence of a transition is still an open point.
Here, we have been able to reach a regime where $\te$ has no
noticeable time dependence. Interestingly enough, it coincides with
the full-aging regime.  At the lowest temperature, the $\te$ measured
over a time window of $3$ orders of magnitude approaches a finite
value, different from the equilibrium temperature. This seems little
compatible with a critical slowing down ($\te \to T$) or the growth of
a crystal phase ($\te \to \infty$) and favors rather the phase
transition scenario.  Our result suggests that the relevant
information to an understanding of aging in structural glasees has to
be looked for in a regime that so far had not been investigated,
either in experiments or in simulations.

We thank A.~Cavagna, B.~Coluzzi, L.~A.~Fern\'andez and F.~Sciortino
for discussions. Simulations used the full RTN3 cluster of U.~Zaragoza
for 6 months.  V.M.-M. is {\em Ram\'on y Cajal} research fellow, and
was partially supported by MCyT (Spain) (FPA2001-1813 and
FPA2000-0956). P.V. was supported by the ECHP programme, contract
HPRN-CT-2002-00307, {\em DYGLAGEMEM}.


\begin{thebibliography}{99}

\bibitem{liquids} C. A. Angell, Science {\bf 267,} 1924 (1995).

\bibitem{Struick78} L. C. E. Struick, {\em Physical aging in amorphous
polymers and others materials}, (Elsevier, Houston, 1978).

\bibitem{colloids} D. Bonn, H. Tanaka, G. Wegdam, H. Kellay, and
J. Meunier, Europhys. Lett. {\bf 45,} 52 (1998).

\bibitem{sg} M. M\'ezard, G. Parisi, and M. A. Virasoro, {\em Spin Glass
Theory and Beyond,} World Scientific, Singapore (1987).

\bibitem{DYNBOOK} J.-P. Bouchaud, L. F. Cugliandolo, J. Kurchan, and
M. M\'ezard in: {\em Spin-glasses and random-fields,} edited by
A. P.Young (World Scientific, Singapore, 1998).

\bibitem{AGINGKURCHAN} J. Kurchan, Comptes Rendus de Physique de
l'Academie des Sciences {\bf IV,} 239 (2001).

\bibitem{Vincent97} E. Vincent, J. Hammann, M. Ocio, J.-P. Bouchaud,
and L. F. Cugliandolo in: {\em Complex behaviour of glassy systems},
edited by M. Rub\'\i\ and C. P\'erez-Vicente (Springer-Verlag, New
York, 1997).

\bibitem{agingsg} G.F.Rodriguez, G. G. Kenning, and R. Orbach,
Phys. Rev. Lett. {\bf 91,} 037203 (2003); S. Jimenez,
V. Mart\'\i{}n-Mayor, G. Parisi, and A. Taranc\'on, J. Phys. A:
Math. and Gen. {\bf 36,} 10755 (2003).

\bibitem{Bouchaud99} J.-P. Bouchaud in: {\em Soft and Fragile Matter:
Nonequilibrium Dynamics, Metastability and Flow,} edited by
M. E. Cates and M. R. Evans (IOP publishing, Bristol, 2000).

\bibitem{colloids-ag} B. Abou, D. Bonn, and J. Meunier, Phys. Rev. E,
{\bf 64} 021510 (2001); A. Knaebel, M. Bellor, J.-P. Munch,
V. Viasnoff, F. Lequeux, and J. L. Harden, Europhys. Lett. {\bf 52},
73 (2000).

\bibitem{Ciliberto03} L.Buisson, L. Bellon, and S. Ciliberto, J. Phys:
Cond. Matt. {\bf 15}, S1163 (2003).

\bibitem{Kob97} W. Kob and J.-L. Barrat, Phys. Rev. Lett. {\bf 78},
4581 (1997).

\bibitem{Nagel98} R. L. Leheny and S. R. Nagel, Phys. Rev. B {\bf 57,}
5154 (1998).

\bibitem{FDTexpliq} T. S. Grigera and N. E. Israeloff,
Phys. Rev. Lett. {\bf 83,} 5038 (1999).

\bibitem{FDTexpnoliq} L. Bellon, S. Ciliberto, and C. Laroche,
Europhys. Lett. {\bf 53}, 511 (2001); L. Bellon and S. Ciliberto,
Physica D {\bf 168,} 325 (2002); D. H\'erisson and M. Ocio,
Phys. Rev. Lett. {\bf 88,} 257202 (2002); L. Buisson, S. Ciliberto,
and A. Garcimart\'\i{}n, Europhys. Lett. {\bf 63,} 603 (2003).

\bibitem{CuKuPe} L. F. Cugliandolo and J. Kurchan, Phys. Rev. Lett. {\bf
71,} 173 (1993); L. F. Cugliandolo, J. Kurchan and L. Peliti,
Phys. Rev. E {\bf 55,} 3898 (1997).

\bibitem{FDTnumliq} G. Parisi, Phys. Rev. Lett. {\bf 79,} 3660 (1997);
J. L. Barrat and W. Kob, Europhys. Lett. 46, 637 (1999); R. Di
Leonardo, L. Angelani, G. Parisi, and G. Ruocco. Phys. Rev. Lett. 84,
6054 (2000).

\bibitem{Sciortino01} F. Sciortino and P. Tartaglia,
Phys. Rev. Lett. {\bf 86,} 107 (2001).

\bibitem{fictive} A. Q. Tool, J. Am. Ceram. Soc. {\bf 29,} 240 (1946);
O. S. Narayanaswamy, {\sl ibid.} {\bf 54,} 491 (1971); C. T. Moyniham,
{\sl ibid.} {\bf 59,} 12 (1976); {\bf 59,} 16 (1976).

\bibitem{Goetze92} W. G\"otze, L. Sj\"ogren, Rep. Prog. Phys. {\bf
55,} 241 (1992).

\bibitem{SWAP} T. S. Grigera and G. Parisi, Phys. Rev. E {\bf 63,}
045102(R) (2001).

\bibitem{ss} J. L. Barrat, J.-N. Roux, and J.-P. Hansen,
Chem. Phys. {\bf 149,} 197 (1990).

\bibitem{Santen99} L. Santen and W. Krauth, Nature {\bf 405}, 550 (2000).

\bibitem{cm} The center of mass (CM) was constrained to be fixed in
order to avoid a spurious fast decay of the correlation due to 
random fluctuations of the CM position.

\bibitem{Jacknife} J. Shao and D.  Tu, {\em The Jackknife and
Bootstrap,} Springer-Verlag, New York (1995).

\bibitem{zero} L. Berthier and J. P. Garrahan J. Chem. Phys. {\bf
119}, 4367 (2003); S. Whitelam, L. Berthier, and J. P. Garraham,
cond-mat/0310207.

\bibitem{tarjus} D. Kivelson, S. A. Kivelson, X. Zhao, Z. Nussinov,
and G. Tarjus, Physica A {\bf 219}, 27 (1995).

\bibitem{trans} M. M\'ezard and G. Parisi, Phys. Rev. Lett. {\bf 82,}
747 (1998); B. Coluzzi, G. Parisi, and P. Verrocchio,
Phys. Rev. Lett. {\bf 84,} 306 (2000).

\bibitem{crystal} A. Cavagna, I. Giardina, and T. S. Grigera,
J. Chem. Phys. {\bf 118,} 6974 (2003).

\bibitem{kob-indep} T. Gleim, W. Kob, and K. Binder,
Phys. Rev. Lett. {\bf 81,} 4404 (1998).

\bibitem{FDTequivalence} E. Marinari, G. Parisi, F. Ricci-Tersenghi,
and J. J. Ruiz-Lorenzo, J. Phys. A: Math. Gen. {\bf 31,} 2611 (1998);
S. Franz, M. M\'ezard, G. Parisi, and L. Peliti, Phys. Rev. Lett. {\bf
81,} 1758 (1998).

\end{thebibliography}
\end{document}